\documentstyle[pre,aps]{revtex}
\draft
\begin{document}

\title{Surface extrapolation length and director structures in confined nematics}
\author{N. Priezjev and Robert A. Pelcovits}
\address{Department of Physics, Brown University, Providence, RI 02912}
\date{\today}
\maketitle

\begin{abstract}

We report the results of Monte Carlo simulations of the Lebwohl--Lasher model of nematic liquid crystals confined to cylindrical cavities with homeotropic anchoring. We show that the ratio of the bulk to surface couplings is not in general equal to the corresponding parameter $K/W$ used in elastic theory (where $K$ is the Frank elastic constant in the one constant approximation and $W$ is the surface anchoring strength). By measuring the temperature dependence of $K/W$ (which is equivalent to the surface extrapolation length) we are able to reconcile the results of our simulations as well as others with the predictions of elastic theory. We find that the rate at which we cool the system from the isotropic to nematic phase plays a crucial role in the development of the final director structure, because of a large free energy barrier separating different director structures as well as the temperature dependence of $K/W$. With a suitably fast cooling rate we are able to keep the system out of a metastable planar state and form an escaped radial structure for large enough systems.

\end{abstract}

\section{Introduction}
\label{sec:Introduction}

The properties of nematic liquid crystals in confined geometries 
continues to be an interesting problem for both 
basic science and technological reasons \cite{Crawford-Zumer}.
Nematics confined to cylindrical cavities with homeotropic
radial boundary conditions can exhibit a variety of nontrivial
structures depending on the competition between bulk elastic and
surface energies.  By minimizing the Frank elastic free energy Cladis
and Kleman \cite{Cladis} and Meyer \cite{Meyer} showed that for a
cylinder with a sufficiently large radius an escaped radial (ER)
configuration will form. This configuration can be thought of as a
planar radial (PR) structure (i.e. a disclination line) which has
``escaped'' along the axis of the cylinder (see Fig.~\ref{fig1}).
Later studies \cite{Crawford} showed the possibility of additional
structures including planar polar (PP) and planar polar with two line
defects (PPLD). In  these latter configurations the director lies in a
plane perpendicular to the axis of the cylinder with a strong
component along a single in-plane direction (see Fig.~\ref{fig2}). In
the PP structure the local director is uniform near the axis of the
cylinder and radial at the boundary. In the PPLD structure there are
two half-integer disclination lines parallel to the cylinder axis.

Kralj and Zumer \cite{Kralj} carried out a very complete numerical stability
analysis of the various nematic structures in a cylindrical geometry
by minimizing the Frank elastic energy (including the saddle-splay
elastic constant $K_{24}$). For most values of the elastic constants they found a phase diagram
with ER and PP structures, with the ER structure stable for large
radii and/or strong anchoring. In particular if the bend and splay elastic constants are equal 
and $K_{24}=0$, then the ER structure should form when $RW/K$ exceeds about 27 \cite{footnote}, where $R$, 
$W$ and $K$ are the cylinder radius (measured in units of intermolecular spacing), surface anchoring strength, 
and Frank elastic constant respectively. However, a PR structure appears if
the twist constant is very large compared with splay. They also found
that the PPLD structure can be stabilized if $K_{24}$ is nearly zero,
R is approximately 100, and the half-integer
defect lines are separated by a distance approximately equal to the diameter of the
cylinder. An analytic study of these director structures was carried out by Burylov \cite{Burylov}
 with similar results.

Monte Carlo (MC) simulations of a cylindrically confined nematic with
radial homeotropic boundary conditions were first performed by Chiccoli {\it et
al.} \cite{Chiccoli} using the Lebwohl--Lasher lattice model
\cite{Lebwohl}.  This model represents each mesogenic molecule or small group
of molecules by a three-dimensional spin vector. These spins are free
to rotate about their centers which are fixed on the sites of a lattice.  Spins whose distance $r$ from the center of
the cylinder is less than $R$, the radius of the cylinder, interact with their nearest-neighbors through the usual bulk
Lebwohl-Lasher pair potential:

\begin{equation}
U_{ij} = - \epsilon_{b} P_{2}(\cos \theta_{ij}) 
\label{bulkenergy}
\end{equation}
where $\theta_{ij}$ is the angle between two spins $i$ and $j$, 
$P_2$ is a second rank Legendre polynomial and $\epsilon_b$ is 
a positive constant for nearest--neighbor sites  and zero otherwise. The homeotropic surface anchoring is introduced via a group of boundary layer spins located at distances $R<r<R_c$ which point in a fixed radial direction.   The interaction of these spins with the spins inside
the cylinder is described by the potential: 

\begin{equation} 
U_{ik} = - \epsilon_{s} P_{2}(\cos\theta_{ik})
\label{surfenergy} 
\end{equation} 
where spin $i$ is located inside the
cylinder and spin $k$ is the nearest--neighbor spin which belongs to
the boundary layer. The coupling $\epsilon_s$ is the surface anchoring
strength. The total potential energy is given by the sum of the two
energies above, summed over all pairs of spins. Note that in the
Lebwohl-Lasher model, because the energy is invariant under a uniform
rotation of all the spins, the bend, splay and twist elastic constants
are equal, and the saddle splay elastic constant $K_{24}$ is
identically zero.

Chiccoli {\it et al.}~studied cylinders of radii 6 and 11 lattice spacings and heights 40, 52, 62, and 82 lattice spacings with periodic boundary conditions along the cylinder axis. The spins were placed on the sites of a simple cubic lattice and the director structures were studied at a temperature of 0.6, measured in units of $\epsilon_b$. Various values of the ratio of the surface 
anchoring strength $\epsilon_s$ to the bulk nearest-neighbor coupling $\epsilon_b$ were 
considered, the largest being unity. In all cases, a planar structure
was observed; however, these authors did not determine whether this structure is PP
or PPLD.  

Subsequently Smondyrev and Pelcovits \cite{Smondyrev:99} 
carried out Monte Carlo simulations on much larger Lebwohl--Lasher systems up to radii $R=160$, with a height 
of 16, again using a simple cubic lattice. For low values of the radius or anchoring strength they found a PPLD structure, whereas for $RW/K=160, \epsilon_s/\epsilon_b=1$, and a temperature of 0.9, they obtained a metastable ER configuration. But this metastable structure eventually collapsed to a planar state, apparently in contradiction with elastic theory which \cite{Kralj} predicts that the ER structure should be stable when $RW/K$ exceeds approximately 27. If we assume 
(incorrectly, as we show in Sec.~\ref{sec:Extrapolation length}) that the ratio $\epsilon_s /\epsilon_b$ equals the corresponding 
Frank elastic energy ratio $W/K$, then Smondyrev and Pelcovits were simulating a system with $RW/K =160$, far in excess of the critical value of 27.

Brada{\v c} {\it et al.} \cite{Bradac} carried out molecular dynamics simulations on cylindrically confined nematics interacting with a modified induced--dipole--induced--dipole interaction. For systems quenched from the isotropic phase to temperatures deep in the nematic phase, a phase diagram was obtained that is in reasonable agreement with the predictions of elastic theory. For the largest systems studied (radii of 16 lattice spacings) the PP, PPLD and ER structures were observed with increasing values of the surface interaction. For the case of equal elastic constants (i.e. like the Lebwohl--Lasher model) the ER structure appears when the ratio of the surface to bulk interaction energies is 0.2. However, these simulations were carried out for only 10000 time steps leaving open the possibility that the ER structure ultimately collapses. Also, these simulations were carried out at apparently lower temperatures than those used in the Monte Carlo simulations.

In this paper we report the results of MC simulations of the confined Lebwohl--Lasher model which address the question of why the ER structure was not observed in previous MC simulations.  We find that the ratio of the bulk--to--surface couplings in the Lebwohl--Lasher model, $\epsilon_b / \epsilon_s$, is {\it not} equal to $K/W$, except at very low temperatures. The latter quantity is equivalent \cite{deGennes:93} to the surface extrapolation length $b$. We find that there is significant dependence of $b$ on temperature in the Lebwohl--Lasher model; in particular $b$ grows as the NI transition temperature is approached from below (a feature observed in experimental studies of confined nematics \cite{Mertelj,Kobayashi}).  Thus, identifying the parameter $R \epsilon_s /\epsilon_b$ with $RW/K \equiv R/b$ is in general incorrect. Except at very low temperatures $RW/K$ will be less than $R \epsilon_s /\epsilon_b$ (substantially so near the NI transition). Comparing the results of simulations with the predictions of elastic theory then requires consideration of this difference. 

We also find, for large enough systems, that the rate at which the system is cooled from the isotropic to the nematic phase determines which director structure is ultimately observed in a simulation. Because $b$ is large near the NI transition, a slow cooling rate will first yield a PPLD structure in this temperature range, in agreement with elastic theory which predicts that the PPLD is stable for small values of $R/b$. However, because of the large free energy barrier between the ER and PPLD structures, continued slow cooling to lower temperatures (where $b$ is small and elastic theory predicts the appearance of the ER structure for sufficiently large $R$) will not yield the expected ER structure; the system becomes trapped in the metastable PPLD state. On the other hand, a relatively rapid cooling from the isotropic phase to low temperatures bypasses the PPLD state and allows the ER structure to form. To our knowledge, this is the first time a stable ER structure has been seen in a MC simulation of confined nematics. 

Our paper is organized as follows. In the Sec.~\ref{sec:Extrapolation length} we describe our measurements of the surface extrapolation length and its temperature dependence. In Sec.~\ref{sec:simulations} we report on simulations of a cylindrically confined nematic, exploring the effect of different cooling rates on the formation of the ER and PPLD structures. We offer some concluding remarks in the final section.

\section{Surface Extrapolation length}
\label{sec:Extrapolation length}

To measure the temperature dependence of the surface extrapolation length $b$, we consider a modification of the discussion presented in Ref. \cite{deGennes:93}, where this length is defined by examining the director profile in a nematic confined between two plates which impose a twist distortion that is uniform across the sample except near the plates. The twist geometry is appropriate for studying planar anchoring at a surface. As we are interested in homeotropic alignment we consider instead a splay geometry.  As shown in Ref. \cite{deGennes:93} the extrapolation length can be defined by extrapolating the constant slope portion of the director profile corresponding to the uniform twist or splay distortion; see Fig.~\ref{extrap}. By minimizing the total elastic energy consisting of the bulk Frank energy in the one--elastic constant approximation and the surface energy, it was shown in Ref. \cite{deGennes:93} that the length $b$ so defined is related to the Frank elastic constant $K$ and the surface energy $W$ by $b=K/W$. 

We measured $b$ defined in the above fashion in the Lebwohl--Lasher model on a cubic lattice of linear size 40. On the plane $z=40$ we introduced the homeotropic surface interaction, Eq.~(\ref{surfenergy}), with boundary spins aligned along the $z$ axis. Spins on the opposite wall, $z=0$, were rigidly aligned parallel to the $x$ axis. These spins interact with their interior neighbors via the bulk interaction, Eq.~(\ref{bulkenergy}). Periodic boundary conditions were imposed on the remaining four sides of the lattice. 
Starting in the isotropic phase the system was cooled down in 
temperature steps of 0.05. We equilibrated at least $200000$ MC steps at each value of the temperature. We display results in Fig. \ref{extrap} for three values of temperature ranging from deep in the nematic phase to near the NI transition.  For each value of $z$ we plot the average value of the spin's angle with the $z$ axis (averaging over all sites in the $x-y$ plane). We note that $b$ appears to diverge near the NI transition temperature in agreement with experimental results \cite{Mertelj,Kobayashi}. Additional data for $b$ as a function of $T$ and $\epsilon_s$ is shown in Table \ref{table}. As expected, $b$ increases with decreasing surface coupling.

The strong dependence of $b$ on temperature indicates that $W$, the surface energy, has a different dependence on temperature than the Frank elastic constant $K$. The latter was measured in the Lebwohl--Lasher model by Cleaver and Allen \cite{Cleaver:91}, who found it to be approximately proportional to $S^2$, where $S$ is the nematic order parameter. Experimental measurements of $b$ \cite{Mertelj} find that it diverges as $(T_{NI}-T)^{-1}$, where $T_{NI}$ is the NI transition temperature, which suggests that $W$ is proportional to $S^4$. Our data is not sufficiently robust to determine the dependence of $W$ on $S$.

The temperature dependence of $b$ helps to explain the apparent discrepancy between the results of previous simulations \cite{Chiccoli,Smondyrev:99} and elastic theory \cite{Crawford}, summarized in Sec.~\ref{sec:Introduction}. Even at a relatively low temperature $T=0.6$ as was used in Ref. \cite{Chiccoli}, $b$ is of order 2, and thus the elastic theory parameter $RW/K=R/b$ that determines the relative stability of the various director structures is approximately half the value of the corresponding Lebwohl--Lasher ``bare'' parameter, $R\epsilon_s/\epsilon_b$. Ref. \cite{Smondyrev:99} reported on simulations carried out at $T=0.9$ with $\epsilon_s/\epsilon_b=1.0$, where the Lebwohl--Lasher parameter is approximately 6 times greater than $RW/K$. Thus the simulations carried out in this latter work at a value of $R=160$, in reality correspond to $RW/K \approx 27$, i.e., approximately the threshold value for the formation of the ER structure. On the other hand, the simulations of Ref. \cite{Bradac} were carried out deep in the nematic phase where $RW/K$ and $R\epsilon_s/\epsilon_b$ are nearly equal.

\section{Stability of the ER and PPLD structures}
\label{sec:simulations}

With the insight gained in the previous section regarding the proper comparison of simulation results with elastic theory, we carried out additional MC simulations on the Lebwohl--Lasher model confined to a cylinder. Previous simulation studies of nematics confined to cylinders \cite{Chiccoli,Smondyrev:99,Bradac} were done on a cubic lattice. However,
this choice of lattice biases the position of the two disclination lines in the PPLD
structure; in particular they always appear on one of the two crystalline axes perpendicular to the long axis of the cylinder. In addition approximately half of the spins on the perimeter of the cylinder have {\it two} nearest--neighbor spins on the fixed boundary layer (see Eq. (\ref{surfenergy})), whereas the other half have only one such nearest--neighbor, resulting in a nonuniform coupling of the nematic to the cylindrical surface.

To eliminate this nonuniformity we worked instead with a honeycomb lattice.
With this choice nearly all spins on the perimeter of the cylinder have only one
nearest--neighbor in the boundary layer.
While the honeycomb lattice does not have continuous rotational symmetry about the cylinder axis, we found that its higher symmetry about this axis (six--fold as compared to four--fold for the cubic lattice) led to the appearance of the disclination lines in the PPLD structure on axes not necessarily along the crystal lattice directions. 
The honeycomb lattice is bipartite allowing us to implement a 
standard ``checkerboard'' algorithm \cite{Binder:95}, an efficient choice for vector processors. We divided the system into two interconnected 
sublattices and then alternately updated spins on each.

As in Ref. \cite{Smondyrev:99} we determined the nature of the resulting director structures by computing 
an escaped order parameter, which measures the degree of spin tilt
away from the plane perpendicular to the cylinder axis, and is a sensitive measure of the presence of the ER structure. It is given
by:
 \begin{equation} P_{2}^{esc} = (1/N)\sum P_{2}({\bf \hat
u_{i}}\cdot {\bf \hat z}) \end{equation}
where the cylinder axis is
along the $z$ direction. The escaped order parameter is calculated as
a function of the distance from the center of the cylinder, and the sum is over spins in a thin annulus with the same height as the cylinder.

We explored the stability of the PPLD and ER structures for systems of two different radii, 80 and 120, both of height 16, with $\epsilon_s/\epsilon_b =0.9$. In Table \ref{table2} we tabulate the values of $R/b$ for these two radii using the data for $b$ from Table \ref{table}. Using the table we see that elastic theory predicts that the PPLD structure should be stable for temperatures greater than 0.6 and 0.8, for $R=80$ and 120, respectively, while the ER structure should be stable at temperatures below these values. Note that elastic theory by its nature is unable to predict the transition temperature separating the PPLD and ER structures from the isotropic phase. Beginning in the isotropic phase at $T=1.4$ we cooled the system of $R=80$, in temperature steps of 0.1 equilibrating 
$100000$ MC steps at each temperature. At $T=0.9$ a stable PPLD structure was formed. We checked the stability of this structure up to 500000 MC steps. Upon further cooling down to $T=0.5$  (again equilibrating 100000 MC steps at each temperature) the system remained trapped in the PPLD structure even though at this lower temperature $b=1.1$ and $R/b = 72.7$, well above the threshold predicted by elastic theory for the formation of the ER structure. We found similar behavior irrespective of the cooling rate used, including instantaneous quenches directly from the isotropic phase. Conversely,  if we began at $T=0.8$ with an  ER configuration constructed by hand using the mathematical form derived in Refs. \cite{Cladis,Meyer}, we found this state to be stable even after 500000 MC steps. Yet for this value of $T$ the PPLD state should be stable ($R/b =20.0$). These results suggest that there is a large free energy barrier separating the PPLD and ER structures, and the system can be trapped in either state.

We found somewhat different results for the larger radius system, $R=120$. We cooled this system down from $T=1.4$ again in temperature steps of 0.1 equilibrating 300000 MC steps at each value of temperature. As expected a stable (up to 500000 MC steps) PPLD structure was formed at $T=0.9$ ($R/b \simeq 20 < 27$). Upon further cooling the system remained trapped in the PPLD state, just as in the case of the smaller system. However, if the same system were equilibrated $100000$ MC steps at each temperature step as it was cooled from $T=1.4$ to $T=0.8$, then an ER structure was formed at the latter temperature and remained stable up to 500000 MC steps. In this case the more rapid rate of cooling allowed us to bypass the incipient PPLD structure at $T=0.9$ and reach the ER structure which should be, according to elastic theory, the globally stable one at $T=0.8$ (where $b \simeq 4$, and $R/b \simeq 30$). We suspect that our ability to form an ER structure for this system of larger radius is due to the smaller temperature range ($0.85 \pm 0.05$) where the PPLD structure is stable compared with the case of $R=80$, where the PPLD structure, according to elastic theory and Table \ref{table2}, should be stable from $T \simeq 0.9$ down to $T\simeq 0.7$.

\section{Conclusions}
\label{sec:conclusions}

In this paper we have addressed the question of reconciling the predictions of elastic theory with the results of numerical simulations for nematics confined to cylindrical cavities with homeotropic boundary conditions. Our main result is that care must be taken in comparing elastic theory with numerical simulations because of the strong temperature dependence of the ratio $K/W$, the surface extrapolation length. Except at very low temperatures, this ratio is much {\it larger} than the corresponding ``bare'' ratio of bulk and surface energies in a numerically simulated model. When this difference is taken into account, the results of numerical simulations are consistent with the predictions of elastic theory. In particular, the ER structure forms for values of $RW/K$ above the critical value of 27 predicted by elastic theory, providing that the metastable PPLD structure can be avoided. Because $K/W$ is very large near the NI transition, and thus $RW/K$ is relatively small, the system will readily form a PPLD structure if it is cooled slowly from the isotropic phase. However, with rapid cooling and for large enough systems where the PPLD phase is limited in its temperature range, this state can be bypassed and a stable ER structure formed at lower temperatures where $RW/K$ is sufficiently large. 

\section*{Acknowledgments}

We thank Professor G.~P. Crawford for helpful discussions and Dr. G.~B. Loriot 
for computational assistance. This work was supported by the National 
Science Foundation under grant No. DMR98-73849. Computational work in support 
of this research was performed at Brown University's Theoretical Physics 
Computing Facility. 

\bibliographystyle{prsty}

\begin{thebibliography}{150}

\bibitem{Crawford-Zumer} {\em Liquid crystals in complex geometries}, edited by G.~P. Crawford and S. Zumer, (Taylor \& Francis, London, 1996).


\bibitem{Cladis} P.~E. Cladis and M. Kleman, J. Phys.(Paris), {\bf 33}, 591 (1972). 

\bibitem{Meyer} R.~B. Meyer,  Phil. Mag. {\bf 27}, 405 (1973). 

\bibitem{Crawford} D.~W. Allender, G.~P. Crawford, and J.~W. Doane, Phys. Rev. Lett. {\bf 67,} 1442 (1991); G.~P. Crawford, D.~W. Allender, and J.~W. Doane, Phys. Rev. A {\bf 45,} 8693 (1992); S. Kralj and S. {\v Z}umer, Phys. Rev. A {\bf 45,} 2461 (1992); S. Kralj and S. {\v Z}umer, Liq. Cryst. {\bf 15,} 521 (1993).

\bibitem{Kralj} S. Kralj and S. {\v Z}umer, Phys. Rev. E {\bf 51,} 366 (1995).

\bibitem{footnote} In the one constant approximation this critical value of $RW/K$ separating the ER 
and PP structures can also be obtained analytically; see Ref. \cite{Crawford}. 

\bibitem {Burylov} S.~V. Burylov, So. Phys. JETP {\bf 85,} 873 (1997).

\bibitem{Chiccoli} C. Chiccoli, P. Pasini, F. Semeira, E. Berggren, and C. Zannoni, Mol. Cryst. Liq. Cryst., {\bf 290}, 237 (1996) 

\bibitem{Lebwohl} P.~A. Lebwohl and G. Lasher,  Phys. Rev. A {\bf 47}, 4780 (1972).



\bibitem{Smondyrev:99} A.~M. Smondyrev and R.~A. Pelcovits, Liq. Cryst. {\bf 26}, 235-240 (1999).


\bibitem{Bradac}Z. Brada{\v c}, S. Kralj, and S. {\v Z}umer, Phys. Rev. E {\bf 58}, 7447 (1998).


\bibitem{deGennes:93} 
P. {\lowercase{d}}e~Gennes and J. Prost, {\em The Physics of Liquid Crystals}
  (Clarendon Press, Oxford, 1993).

\bibitem{Mertelj}A. Mertelj and M. {\v C}opi{\v c}, Phys. Rev. Lett. {\bf 81}, 5944 (1998).

\bibitem{Kobayashi}D.-S. Seo, Y. Limura, and S. Kobayashi, Appl. Phys. Lett. {\bf 61}, 234 (1992); T. Sugiyama, S. Kuniyasu, and S. Kobayashi, Mol. Cryst. Liq. Cryst. {\bf 238}, 1 (1994).


\bibitem{Cleaver:91} D.~J. Cleaver and M.~P. Allen, Phys. Rev. A {\bf 43}, 1918 (1991)  

\bibitem{Binder:95} D.~P. Landau, in {\em The Monte Carlo Method in Condensed Matter Physics}, edited by K. Binder, (Springer, Berlin, 1995), second edition.


\end{thebibliography}

\narrowtext
\begin{table}
\caption{Temperature dependence of the extrapolation length $b$ for two different values of the surface coupling $\epsilon_s$. The bulk coupling $\epsilon_b$ is unity in both cases. The system is a cubic lattice of linear size 40.}
\begin{center}
\begin{tabular}{ccc}
\hline
$T$ & $\epsilon_{s}=0.45$ & $\epsilon_{s}=0.9$\\
\hline
1.0 & 10.3 & 9.5\\
0.9 & 7.5 & 6.1\\
0.8 & 6.0 & 4.0\\
0.7 & 4.1 & 3.1\\
0.6 & 2.0 & 1.8\\
0.5 & 1.1 & 1.1\\
\hline
\end{tabular} \end{center} \label{table}\end{table}

\begin{table}
\caption{Temperature dependence of $R/b$ for cylinders of radii 80 and 120 with $\epsilon_s/\epsilon_b =0.9$. The surface extrapolation length $b$ is given in Table \ref{table}. According to elastic theory the PPLD structure should be stable for $R/b$ less than approximately 27, while the ER structure should be stable for larger values of $R/b$.}
\begin{center}
\begin{tabular}{ccc}
\hline
$T$ & $R/b, R=80$ & $R/b, R=120$\\
\hline
1.0 & 8.4 & 12.6\\
0.9 & 14.3 & 19.8\\
0.8 & 20.0 & 30.0\\
0.7 & 25.8 & 38.7\\
0.6 & 44.4 & 66.7\\
0.5 & 72.7 & 109.1\\
\hline
\end{tabular} \end{center} \label{table2}\end{table}

\begin{figure}

\caption{Schematic director pattern in the (a) escaped radial (ER), and (b) planar radial (PR) configurations. The cylinder axis is vertical in (a) and perpendicular to the page in (b).} 
\label{fig1}

\end{figure}

\begin{figure}

\caption{Schematic director pattern in the (a) planar polar (PP), and (b) planar polar with line defects (PPLD) configurations. The cylinder axis is perpendicular to the page.} \label{fig2} \end{figure}

\begin{figure}

\caption{Measurement of the extrapolation length $b$ in a Lebwohl--Lasher model of linear size 40 for three representative temperatures: a) $T= 0.5$, b) $T=0.85$, and c) $T= 1.05$, (the NI transition is approximately 1.1). In each plot the average value of $\theta$, the angle made by the spins with $z$ axis is shown as a function of $z$. The spins at $z=0$ are rigidly aligned parallel to the $x$ axis, while the spins at $z=40$ are coupled homeotropically to the surface with $\epsilon_s=0.9$. Following \protect\cite{deGennes:93} we define the extrapolation length $b$ as shown in the figures. The solid straight lines are the best fits (to the eye) of the bulk splay distortion imposed by the boundary conditions. The numerical values of $b$ for these three temperatures are 1.1, 5.2 and 16.3 respectively. Values for other temperatures and a smaller value of the surface coupling are given in Table \ref{table}. } \label{extrap} \end{figure}

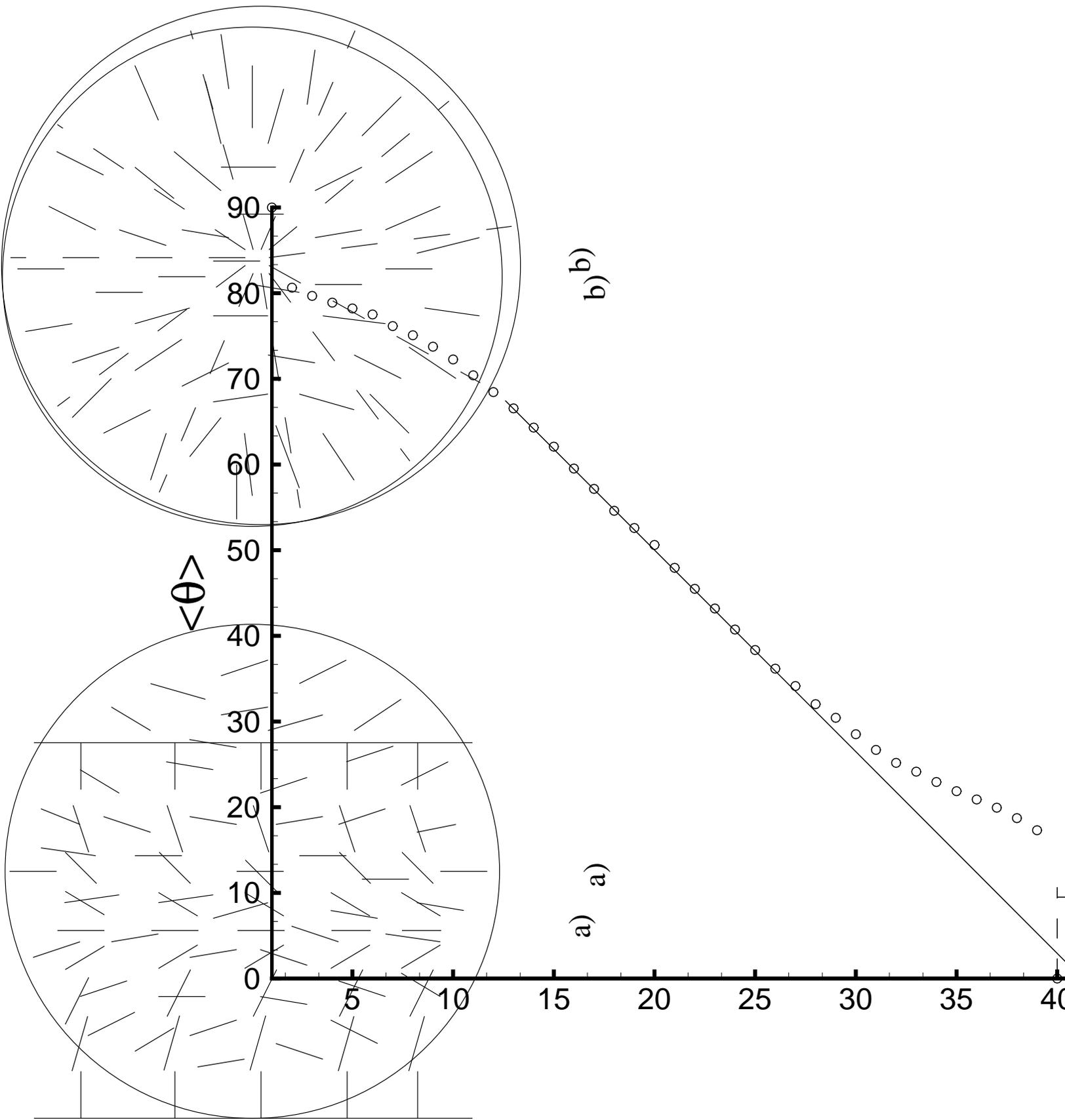
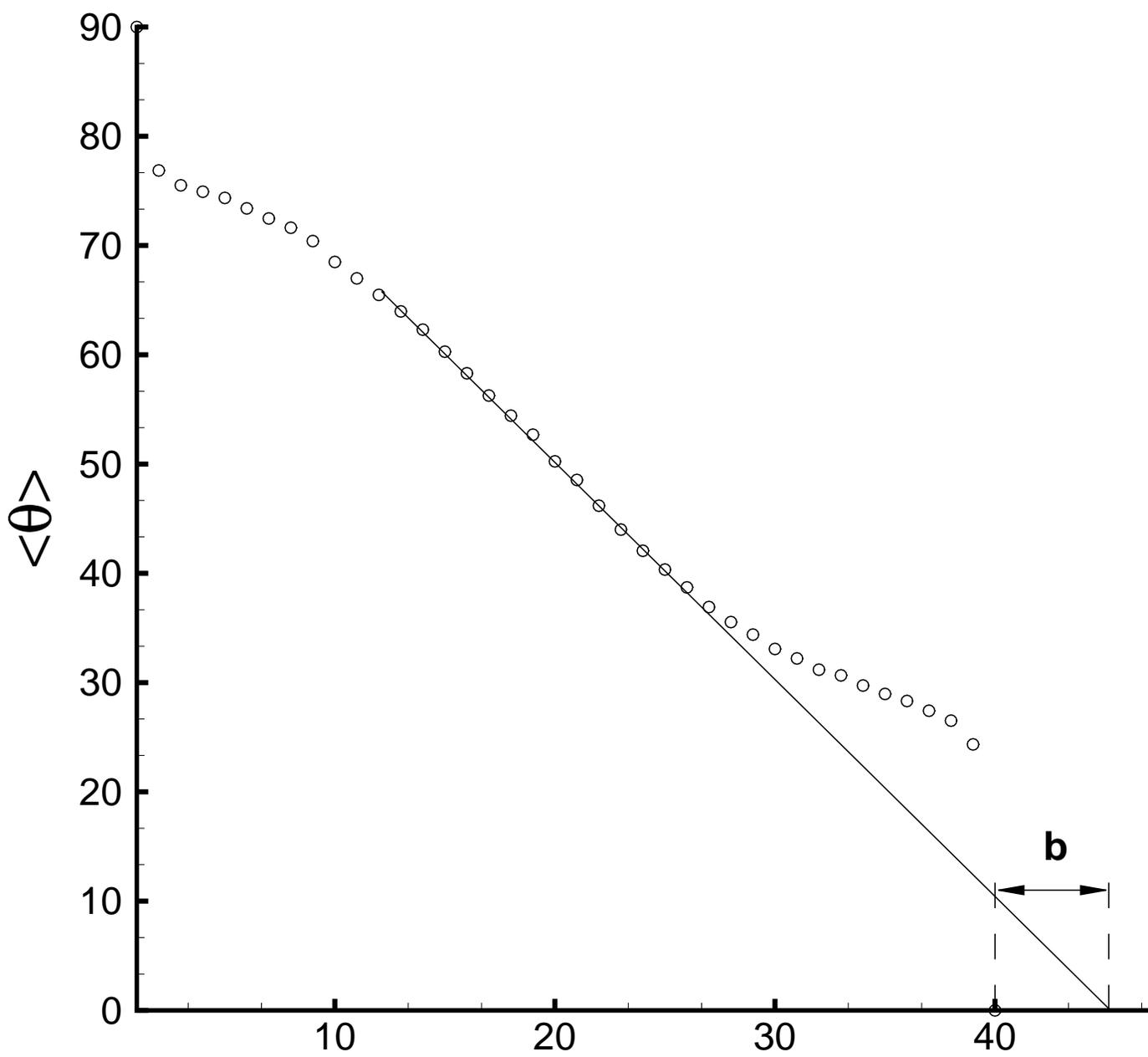
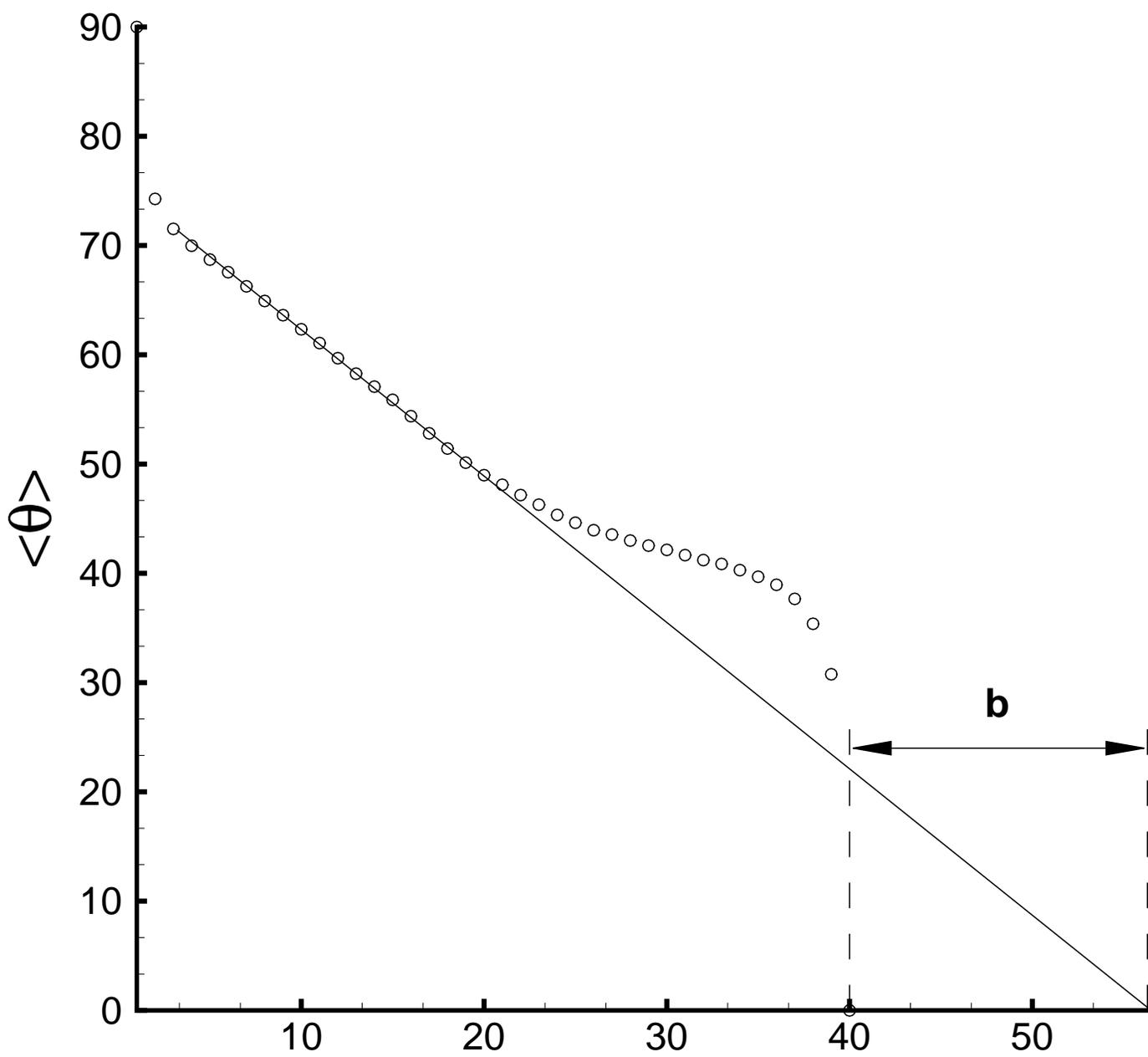
\end{document}